# Structure vs. Language: Investigating the Multi-factors of Asymmetric Opinions on Online Social Interrelationship with a Case Study

Bo Wang
Tianjin University
bo_wang@tju.edu.cn

Yingjun Sun
Tianjin University
380261014@qq.com

Yuan Wang
Tianjin University
939931891@qq.com

## ABSTRACT

Though current researches often study the properties of online social relationship from an objective view, we also need to understand individuals' subjective opinions on their interrelationships in social computing studies. Inspired by the theories from sociolinguistics, the latest work indicates that interactive language can reveal individuals' asymmetric opinions on their interrelationship. In this work, in order to explain the opinions' asymmetry on interrelationship with more latent factors, we extend the investigation from single relationship to the structural context in online social network. We analyze the correlation between interactive language features and the structural context of interrelationships. The structural context of vertex, edges and triangles in social network are considered. With statistical analysis on Enron email dataset, we find that individuals' opinions (measured by interactive language features) on their interrelationship are related to some of their important structural context in social network. This result can help us to understand and measure the individuals' opinions on their interrelationship with more intrinsic information.

## 1. INTRODUCTION

It's an essential problem in social computing to recognize the nature of the social interrelationships between individuals. The model of interrelationship is fundamental to many related studies such as community discovery, influence analysis, link predication and recommendation [2]. There are two primary options to determine the nature of an interrelationship: The first one is to model the interrelationship from an objective view which supposes the properties of interrelationships can be investigated independently of the subjective opinions of participant individuals. The second option supposes the properties of interrelationships are subjective which should be determined by investigating each individual's opinion on their interrelationships respectively. The interrelationship model based on these two strategies can be quite different. For an interrelationship, the objective principle often assign a single value to each property, which is symmetric to both sides of the interrelationship. While the subjective principle naturally leads to an asymmetric measurement of interrelationship' properties because the two individuals may have different opinions on their interrelationship from their own side.

The subjective asymmetric model of interrelationship is more informative and sensitive than the objective relationship model. To perform asymmetric measurement of interrelationship's properties, we need some features which can reflect the subjective opinions of individuals. Researches on signed social networks suggest to predicate the properties' value for each individual according to the network they are embedded in. Linguistic analysis suggests another path forward to leverage textual features to predict individual's opinion.

Intuitively, to describe the individuals' opinions on their interrelationship, the features of their interactive language should be more capable than the features of their social relationships. However, the structural features of their social relationship may still be a latent cause which influence their asymmetric opinions on interrelationship, and can also be helpful to understand and measure individuals' asymmetric opinions indirectly.

In this work, we investigate the relationship between individuals' opinions on their interrelationship and their structural characteristics in social network. To do this, we try to investigate the correlation between the typical language features of interactive language and the typical structural features of social network.

### 1.1 Language and Network Features in Social Network Analysis

Along with the springing up of more and more researches on the natural language processing and the signed social networks, researchers try to combine these two technologies to investigate the nature of the social relationship in recent studies.

Through measuring the ability of the feature sets of social behavioral and textual information to determine the different types of relationships, Adalı et al. [1] drew the conclusion that these two kind of information were practically equivalent between pairs of individuals' interaction in social media. Approach of Pang and Lee [10] extended the model of text corpus-based statistical learning approach proposed by Bramsen et al. [4].Their improvement was inspired by an assumption of homophily, i.e., certain social relationships correlate with agreement on certain topics. Thomas et al. [12] predicted voting patterns by using party affiliation and what had mentioned in speeches, and Tan et al. [13] predicted attitudes about political and social events by utilizing Twitter follows and comments in the same way. The work of West et al. [15] is an important enlightenment of our work which developed a model which combines textual and social-network information to jointly predict the polarity of person-to-person evaluations in the signed social network.

### 1.2 Asymmetric Measurement of Social Interrelationship

Though symmetric model is most widely studied in social relationship, there are also some latest studies analyzed social relationships directionally or asymmetrically in a sense. For directed relationships, Leskovec et al. [9] first considered an explicit formulation of the sign prediction problem. Their prediction methods are based on the theory of social balance and status. Bach et al. [3] and Huang et al. [7] framed sign prediction as a hinge-loss Markov random field, a type of probabilistic graphical model introduced by Broecheler et al. [5]. West et al. [15] developed a model that synthesizes textual and social-network information to jointly predict the polarity of person-to-person evaluations. Wang et al. [14] investigated the subjective asymmetry of interrelationship with interactive language features.

### 1.3 Our Approaches

In previous work, researchers proposed to model the two individuals' asymmetric opinions on their interrelationship with interactive language features. However, current studies do not explore the intrinsic causes of the opinions' asymmetry on interrelationships sufficiently, and does not make full use of the topology information from the social network to explain the opinions on the interrelationships. In this paper, we focus on the factors that cause individuals' asymmetric opinions on interrelationships.

If individuals' opinions only depend on their own subjective ideas, then we can only measure their opinions with subjective features e.g., the features of interactive language. Intuitively, an individual's opinion on one of his social relationships is not only determined by his own ideas, but also affected by other relationships or individuals related to him in the social network. Therefore, in this paper, we try to put a single interrelationship into social network context in order to investigate the correlation between the individuals' opinions and social network features. As in [14], opinions will be characterized by the features of interactive language. This investigation will tell us whether the social network characteristics are intrinsic causes of individuals' opinions on their interrelationship, and shall we introduce topology features into the subjective model of social relationship.

In section 2, we introduce how to characterize the asymmetric opinions on interrelationship with interactive language features. In section 3, we investigate the correlation between structural factors and asymmetric opinions. Finally we introduce the experiments in section 4 and summarize the conclusions in section 5.

## 2. CHARACTERIZE ASYMMETRIC OPINIONS ON INTERRELATIONSHIP WITH INTERACTIVE LANGUAGE FEATURES

In current studies, most of the researches aim at investigating the strength or type (e.g., positive & negative) of the social relationship between two individuals from an objective view by measuring the objective features, e.g., the frequency of the communication between participants and the embeddedness of interrelationship. On this basis, some researches utilize the directed objective features, e.g., the bidirectional communication frequency between two participants. However, it's still difficult to describe subjective opinions accurately using directed objective features only. Therefore, it's a new trend to combine subjective and objective features to model social relationships, especially from subjective view.

Along with the researches on subjective features, researchers find that interactive language is a good choice to obtain subjective features of interrelationship, because the interactive language not only related to the properties of interrelationship closely, but also is very capable to describe individuals' opinions. In this section, we'll compare the objective and subjective features of the interrelationship and then utilize several typical subjective features of interactive language to measure the asymmetric opinions on interrelationship following the method in [14].

### 2.1 Objective Features of Interrelationship

The most widely used objective feature of social relationship is the topology features of social network. Among social network features, the embeddedness of interrelationships and the communication frequency between the participants are two most popular objective features to measure social relationships.

The embeddedness is a feature which can measure the strength of an interrelationship with the coincidence degree of two individuals' interrelationships, i.e., the count of the common 'friends' of the two participants. Embeddedness has been proved to be an effective objective feature which describes the strength of interrelationships, but it is not designed to be directed.

The frequency of certain type of communication between the two participants are also used to represent the strength of an interrelationship in many studies. The frequency of phone calls, emails and mutual comments on social media are all widely used frequency features. Exactly speaking, communication frequency is not an absolute objective feature, because it is a behavior of the participants which is a subjective decision in a sense. The communication frequency can naturally be measured directionally, e.g., we can measure the bidirectional strength of an interrelationship with bidirectional communication frequency. But, the communication frequency cannot measure subjective opinions directly with very limited information. For example, in some cases, though we have similar email frequency with two colleagues, we may have different attitudes towards the interrelationships with the two colleagues respectively.

The subjective features e.g., linguistic features of interactive language can be used to recognize the participants' opinions on their interrelationship more directly. For the same example, though we have similar email frequency with two colleagues, our different attitudes on two interrelationships still can be revealed by the content of the emails.

### 2.2 Sociolinguistics and Subjective Features

The subjective feature can reveal the participants' opinions on their interrelationship. Based on it, the theory of communicative action [6] in sociolinguistics proposed to reconstruct the concept of relationship with the communicative action instead of the objectivistic information. Thus we can utilize the linguistic structures of communication to understand the social relationships more normatively. Sapir-Whorf hypothesis [11] also supposed that the semantic structure of the language use shapes the ways in which a speaker forms conceptions of the world including social relationships. Individuals' choices of the words in communication were highly related to their opinions on their interrelationship. Therefore we can investigate the asymmetric opinions on social relationships with the interactive language used.

How can we describe an individual's interactive language style in order to describe his opinion on his interrelationship? In sociolinguistics, Holmes [8] proposed four important dimensions to study the language used in social communication:

1. The solidarity-social distance scale: concentrate on the solidarity of the relationships in social communication.

2. The social status scale: concentrate on the relative status of the individuals in social relationship.

3. The formality scale: concentrate on the formality of language that individuals use in different places, topics and relationships.

4. The referential and affective scale: concentrate on referential and affective function of the language that individuals use in social communication.

Among these four dimensions, the first two mainly concern about the features of social interrelationship from both subjective and objective views. The last two aim at the features of interactive languages, which are highly related to the social interrelationship.

Inspired by Holmes' theory, in this work, we use four typical features of interactive language to investigate individuals' opinion on their interrelationships, including frequency, length, fluency and sentiment polarity which indicate quantity, quality and emotion of the interactive language, respectively:

1. The frequency is the number of times of communications within a specified period of time, which can partially reflect one's intentions of the communication on an interrelationship.

2. The length is the number of the words of the interactive language, which is related to one's emphasis degree on an interrelationship.

3. The fluency is a widely used feature in natural language processing which can reflect the formality and quality of interactive language. We can speculate whether an individual treats an interrelationship seriously or not with the quality of his language in communication.

4. The sentiment polarity can indicate the emotion tendency of the interactive language, which tends to have a positive correlation with the personal evaluation on an interrelationship intuitively.

## 2.3 Distinguish the Opinions and Language Habits in Interactive Language Style

Though the features of interactive language are closely related to individuals' opinions on their interrelationship, it is still difficult to understand the opinions accurately with the original value of language features, because in natural language understanding, the true meaning of a sentence is always determined by not only the sentence itself, but also the context.

The context of interactive language is so complex including the occasion of the dialogue, speaker's habit of language use, the sentences in the same discourse and so on. In this work, we focus on the individual's language habit and its influence on the understanding of his interactive language. The meaning of a speaker's language is intuitively related to his language habit. For example, suppose $A$ is a negative people who talks to every people very negatively. We also know that $A$ talks to another person $B$ with negative sentiment polarity score, but $A$ talks to $B$ most friendly compared with the others. In this case, if we want to measure $A$'s opinion on his relationship with $B$ correctly with sentiment polarity score, we need to avoid the interference from $A$'s language habit and use the relative score normalized by $A$'s personal language habit instead of using the original score directly.

In this work, for an individual $I$, to characterize $I$'s opinion more accurately and avoid the interference from $I$'s language habit, we normalize $I$'s each language feature $f$ with $I$'s language habit value $H_f(I)$. $H_f(I)$ is measured by Formula (1), where $f(I, I_i)$ is the $f$'s value of the languages from $I$ to another individual $I_i$. $C$ is the set of all individuals who is in communication with $I$.

$$H_f(I) = \frac{1}{|C|}\sum_{I_i \in C} f(I, I_i) \quad (1)$$

According to $I$'s language habit, we design Formula (2) to calculate $f'(I, I_i)$, which is the normalized value of $f(I, I_i)$:

$$f'(I, I_i) = \frac{f(I, I_i) - H_f(I)}{H_f(I)} \quad (2)$$

## 3. STRUCTURAL FACTORS RELATED TO THE ASYMMETRIC OPINIONS ON SOCIAL INTERRELATIONSHIP

As mentioned in section 1.3, in practical experience, an individual is always not only involved in a single social relationship. His opinion on one interrelationship is often influenced by the social network context he is embedded in. The context here is mainly presented as the structural characteristics, such as other interrelationships and neighbor nodes of this individual.

In this section, we put a single interrelationship into the social network in order to explore the correlation between the interactive language features and the social network features of an interrelationship. We focus on three kinds of typical structural features: node features, relationship features and the features of multiple relationship unit (triadic Closure of relationships).

### 3.1 Degree and Clustering Coefficient of Individuals

First, we focus on the structural features of individuals, i.e., the vertexes in social network. Degree and clustering coefficient are two typical features, which are widely used in social network studies. As for a vertex $A$, the degree of $A$ is defined as the number of vertexes connected with $A$, in this work. A vertex's degree can indicate the range of the social relationships of a person. The clustering coefficient is the quotient of the number of edges between the vertexes connected with $A$ divided by the vertex pairs connected with $A$. The clustering coefficient indicate the probability of any two friends of $A$ are also friends in social network. It's also known as an indicator to measure $A$'s ability to gather friends into a cluster. Assume that the degree of $A$ is $n$ and the number of edges between these $n$ vertexes is $k$. Then the clustering coefficient $C(A)$ of $A$ can be calculate with Formula (3):

$$C(A) = \frac{k}{C_n^2} \quad (3)$$

Why do we choose degree and clustering coefficient as vertex structural features to be investigated in this work? First, they are wildly used in sociology and social computing. What's more, in intuition, the range of one's social relationship and the clustering degree of one's friends may affect one's opinion on his social interrelationships, thus affect the opinion's asymmetry on his interrelationship. To explore this intuition, we carry on correlation analysis between these two vertex features and the asymmetry of individuals' opinion using interactive language features.

### 3.2 Embeddedness of Interrelationships

Second, we focus on the structural feature of a single interrelationship, i.e., a single edge in social network. A typical structural feature of single interrelationship is embeddedness. The embeddedness is the number of common vertexes connected with the two end points of one edge. It means the number of common friends of two persons engaged in an interrelationship. Embeddedness is believed to be a good indicator of the strength of social relationship. In this work, we suppose that embeddedness is also related to the degree of opinions' asymmetry on an interrelationship. This assumption is based on the similar idea of the strength measurement of social relationship: it is widely believed that more common friends indicate make two persons connected to each other more tightly. In this work, we suppose that more common friends can also make two persons' opinions on their interrelationship more symmetric. In this work, this assumption will also be investigated by the correlation analysis between the embeddedness and the asymmetry of individuals' opinions using interactive language features.

## 3.3 The Balance of Interrelationship Triadic Closure

### 3.3.1 The traditional balance theory

Third, we focus on the structural feature of a basic social unit consists of multiple interrelationship, i.e., triadic closure. The triadic closure consists of any three vertexes and edges between them, i.e., any three persons and the interrelationship between them. In traditional balance theory, each social relationship is singed with binary tag '+' or '-', which indicate positive or negative relationship. With binary signs, there are four different signs combination for a triadic closure. The balance theory claims that two of them are balanced and the other two are unbalanced, as shown in Figure 1. The balanced triadic closures are stable while the unbalanced ones tend to change to be balanced.

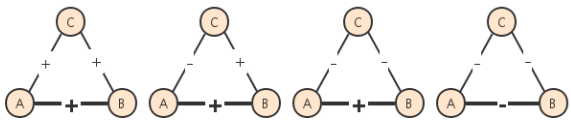

**Figure 1.All chematic diagrams of the triadic closure in the balance theory.The first and third are balanced triangles,while the other two are unbalanced.**

Actually, the traditional balance theory can be explained with the concept of homogeneity. In a triadic closure, the more consistent of the two individuals' views on the third one, the more positive their opinions on their interrelationship is, and vice versa.

### 3.3.2 Our extension of the balance theory

The homogeneity based explanation in previous section extend the traditional balance theory from binary signs to continuous value. Furthermore, this explanation can also extend the traditional balance theory from undirected relationship to directed relaitonship, because it describes the balance with individual's opinion instead of the objective type. This extension makes it possible to investigate the correlation between triangle balance and opinions' asymmetry on interrelationship. In this section, we introduce the details of this extension of traditional balance theory.

First of all, let's analyze the essential idea of the traditional balance theory. The idea is: the polarity of an interrelationship is related to the homogeneity (common friends) of the two participants. To be exact, the homogeneity here indicates the consistency of the views about the third party (person or thing). With this assumption, the traditional balance theory can only predict the binary polarity of an interrelationship between two persons *A* and *B* with the binary polarity of their interrelationship with a third person *C*. It's obvious that in traditional balance theory, we cannot determine the balance of a triadic closure with the individuals' directed opinions instead of the undirected binary signs on interrelationships.

To solve this problem, we extend the traditional balance theory to a directed version keeping the essential principle of homogeneity. In the extension, an individual *A*'s opinion on his interrelationship with another individual *B*, is presented as a continuous value on a directed edge *(A, B)*.When we want to investigate the bidirectional opinions' asymmetry between *A* and *B*, we build an extended triadic closure with four directed edges include *(A, B)*, *(B, A)*, *(A, C)* and *(B, C)* as shown in Figure 2. Instead of the binary sign, we label each directed edge with a real-value in [0, 1] to indicate the opinion of the individual on the start point, e.g., *|AB|* indicates the value of *A*'s opinion on his relationship with *B*.

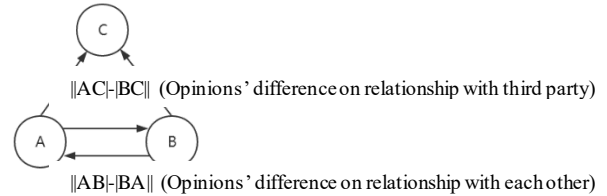

**Figure 2.Extension of the traditional balance theory: determine the balance by comparing the difference between *A* and *B*'s opinions on their interrelationship and the difference between A and B's opinions and their relationship with *C*.**

In this extended triadic closure, we measure the balance of the triadic closure comparing two differences: when the difference between *|AC|* and *|BC|* is smaller than a threshold, we recognize the sign of the opinions' difference on the third party as '-', which means *A* and *B* has similar opinions on their interrelationships with *C*. Otherwise we recognize the sign of the opinions' difference on the third party as '+', which means *A* and *B* have different opinions on their interrelationships with *C*.

In the same way, when the difference between *|AB|* and *|BA|* is smaller than the same threshold, we recognize the sign of the opinions' difference on interrelationship as '-' which means *A* and *B* have similar opinions on their interrelationship. Otherwise we recognize the sign of the opinions' difference on interrelationship as '+' which means *A* and *B* have different opinions on their interrelationship. The threshold here is the average of all bidirectional differences on every interrelationship in the given social network. If the opinions' difference on the third party and the opinions' difference on interrelationship has the same sign, we identify the directed triadic closure as balanced, otherwise, it is unbalanced.

## 4. EXPERIMENTS

In the experiments, we utilized the Enron email dataset to investigate the correlation between the proposed typical social network features and the opinions' asymmetry on interrelationships. The individuals' opinions on interrelationships are characterized with four important language features introduced in section 2.2.

## 4.1 Dataset

In the experiments, we utilized the Enron email dataset of CMU version (May 7, 2015 Version)[1] which contains 0.5M email data exchanged between 151 Enron employees. We choose this dataset because it contains both the interactive language content (email content) and social network (send and receive relationships).

We filtered the email data retaining the users whose email addresses are followed by @enron.com. There are about 0.2M

---

[1] http://www.cs.cmu.edu/~enron/

user pairs who sent/receive emails to each other which indicate the interrelationships between each two users. To make the investigation more reliable, we only kept those interrelationships where at least 15 emails were sent in each direction. The filtered dataset contains 1078 interrelationships between 647 individuals.

### 4.2 Language Features Extraction

For each ordered pair of individuals $<I_i, I_j>$ in the email dataset, we calculated four language features with the content of emails sent from $I_i$ to $I_j$:

1. To calculate the feature "Frequency", assume that the number of emails sent from $I_i$ to $I_j$ is $N$ and the sending date of the first and the last email are $t_1$ and $t_2$, respectively. Then the language feature "Frequency" can be calculated by Formula (4):

$$frequency\_score_{i,j} = \frac{N}{t_2 - t_1} \qquad (4)$$

2. To calculate the feature "Length", assume that the number of emails sent from $I_i$ to $I_j$ is $N$ and the total number of words in these emails is $w$, then the language feature "Length" can be calculated by Formula (5):

$$length\_score_{i,j} = \frac{w}{N} \qquad (5)$$

3. To calculate the feature "Quality", we utilize the SRI language modeling toolkit (SRILM)[2] with Formula (6) to measure the perplexity score which has a negative correlation with the quality of a sentence. In the formula, *prob* is the generating probability of a sentence. Words and oovs are the count of the words and out of vocabulary words in the sentence respectively.

$$perplexity\_score_{i,j} = 10^{(-logprob/(words-oovs+1))} \qquad (6)$$

4. To calculate the feature "Sentiment", we utilize a sentiment dictionary[3] to count the sentiment words. Each positive or negative word is valued 1 or -1, respectively. Assume that there are $S$ sentences in the emails sent from $I_i$ to $I_j$, and the sum of all the scores of sentiment words in $S$ is $W$, then the language feature "Sentiment" can be calculated by Formula (7):

$$Sentiment\_score_{i,j} = \frac{W}{S} \qquad (7)$$

### 4.3 Vertex Factors vs. Language Features

In the first set of experiments, we investigate the correlation between the structural features of the vertexes in social network and the opinions' asymmetry on interrelationships, which is measured by the four language features. The two structural features of the vertexes introduced in section 3.1 will be investigated: degree and clustering coefficient.

#### 4.3.1 Degree vs. Language features

In the first experiment, we firstly calculated the degree of every vertex by counting the number of vertexes connected with it. Then we labeled every directed edges with the value of four language features, respectively. For each vertex $A$, we calculated the average of the asymmetry degree on $A$'s all interrelationships for each language feature. The asymmetry degree on an interrelationship is the degree of bidirectional value differences of a certain language feature. Calculated with Formula (2), $f'(I, I_i)$ is the normalized value of one feature from $I$ to $I_i$, and $C$ is the set of

---

[2] http://www.speech.sri.com/projects/srilm/

[3] http://www.keenage.com/download/sentiment.rar

---

the individuals who is in communication with $I$. Then the average asymmetry $\bar{f}'$ of $I$ is calculated by formula (8).

$$\bar{f}'(I) = \frac{1}{|C|} \sum_{I_i \in C} |f'(I, I_i) - f'(I_i, I)| \qquad (8)$$

Then, we calculated the Pearson Correlation Coefficient between vertexes' degree and their average asymmetry degree on each language feature. The results are shown in the first column in Table 1. Figure 3 illustrates the correlation curves of four language features, where the abscissa is the degree of vertexes, and the ordinate is the average asymmetry degree.

**Table 1. Pearson Correlation Coefficient between structural features of vertexes and vertexes' average asymmetry degree**

| Language Features | Pearson Correlation Coefficient | |
|---|---|---|
| | Degree | Clustering coefficient |
| Frequency | 0.085 | -0.044 |
| Length | -0.650 | -0.378 |
| Quality | -0.109 | 0.189 |
| Sentiment | -0.438 | 0.200 |

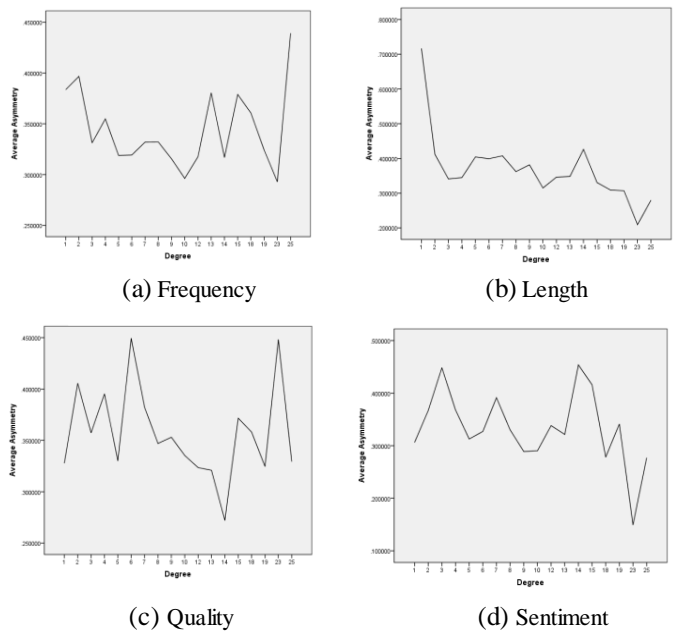

(a) Frequency  (b) Length

(c) Quality  (d) Sentiment

**Figure 3. Curves of correlations between the degree of vertexes (individuals) and the vertexes' average asymmetry degree**

#### 4.3.2 Clustering coefficient vs. Language features

In the second experiment, for each vertex, we calculated the clustering coefficient of every vertex by Formula (3) in section 3.1, as well as the average of the asymmetry degree on $A$'s all interrelationships for each language feature. Then, we calculated the Pearson Correlation Coefficient between vertexes' clustering coefficient and average asymmetry degree on four language features. The results are shown in the second column in Table 1.

Then, we calculated the Pearson Correlation Coefficient between vertexes' clustering coefficient and their average of asymmetry degree on each language feature. The results are shown in the second column in Table 1. Figure 4 illustrates the correlation curves of four language features, where the abscissa is the clustering coefficient of vertexes, and the ordinate is the average asymmetry degree.

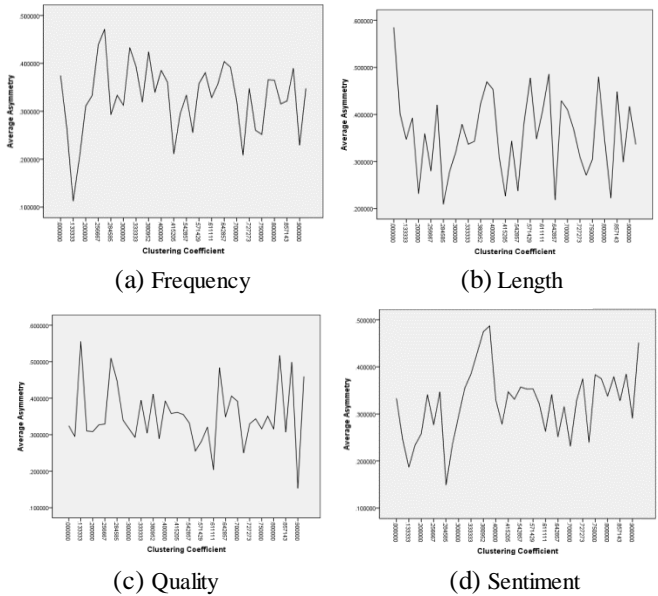

**Figure 4. Curves of correlations between the clustering coefficient of vertexes (individuals) and the vertexes' average asymmetry degree**

### 4.3.3 Observations
We have some initial observations from the statistical results in section 4.3.1 and 4.3.2. In the experiment of vertexes' degree, the asymmetry degree of 'length' and 'sentiment' score relatively have higher negative correlation with the degree of vertexes, and the other two features have no significant correlation. This indicates that a more sociable person, i.e., a person with more friends, may be more active to cater to the partners' opinions in social interrelationships, i.e., have less asymmetry degree on interrelationships, especially reflected on the intension (length) and sentiment tendency (sentiment score).

Though 'length' and 'sentiment' score also have relative higher correlation in the experiment of clustering coefficient, the absolute values are too small, therefore we don't think that there is correlation between the clustering coefficient and the average asymmetry degree of vertexes on any language feature in this case.

Furthermore, the correlation of frequency is the lowest. One possible explanation is that the send/replay relationships between the emails can always lead to similar bidirectional email sending frequency between two individuals. This bidirectional correlation is independent with the opinions or social structure of the individuals.

In general, in this set of experiments, we find that the degree (number of friends) of the individuals in social network may have latent influence on the individuals' opinions and language behaviors on their each interrelationships.

## 4.4 Edge Factors vs. Language Features
In the next experiment, we investigate the correlation between the structural features of single interrelationship in social network and the opinions' asymmetry on the interrelationship, which is also measured by language features.

We calculated the Pearson Correlation Coefficient between the embeddedness of edges and the asymmetry degree between edges' two end points on each language feature. The results are shown in Table 2. Figure 5 illustrates the correlation curves of four language features, where the abscissa is the common friends' number (embeddedness) of edges, and the ordinate is the average asymmetry degree.

**Table 2. Pearson Correlation Coefficient between the embeddedness and average asymmetry degree of edges**

| Language Features | Pearson Correlation Coefficient |
|---|---|
| Frequency | -0.823 |
| Length | -0.979 |
| Quality | -0.678 |
| Sentiment | -0.696 |

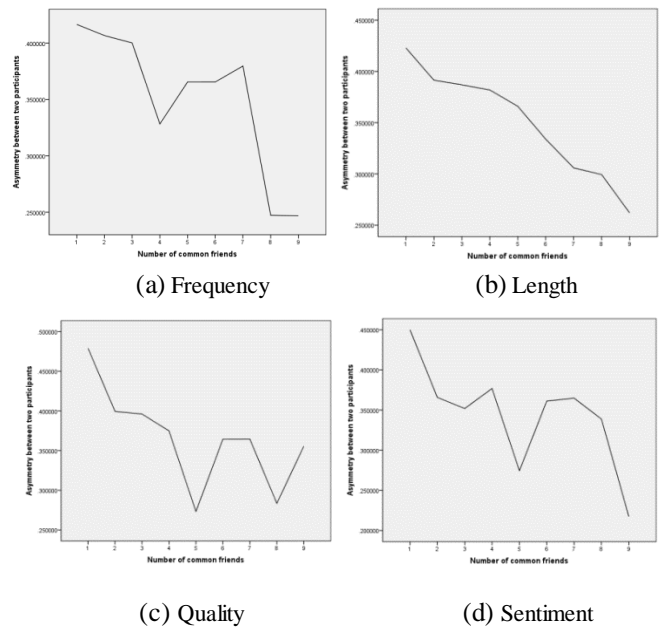

**Figure 5. Curves of correlations between the embeddedness of edges (interrelationships) and the average asymmetry degree on edges**

As shown in the results, we can find that the embeddedness of interrelationship have a high negative correlation with the average asymmetry degree on all four language features. This indicates that if two individuals have more common friends, the asymmetry of their opinions on their interrelationship is weaker, i.e., they have more similar opinions on their interrelationship.

This result can also be an important extension to the traditional triadic closure theory of social relationship. The traditional triadic closure theory states that the strength of a social interrelationship has a positive correlation with the number of common friends. For any two individuals, our results extend the theory to the positive correlation among the number of their common friends, the strength of their interrelationship and the symmetry of their opinions on their interrelationship.

## 4.5 Balance Theory vs. Language Features

### 4.5.1 The verification of traditional balance theory

In the last set of experiments, we firstly verified the balance assumption in traditional balance theory introduced in section 3.3.1 on Enron email data. The balance assumption in tradition balance theory is that in any social network, the signed triadic closures tend to be balanced in general. Since the traditional balance theory is based on undirected edge, we merge every two bidirectional edge pairs between two vertexes into a new undirected edge whose opinion value is the sum of two opinion values on bidirectional edges. Then we set a tunable threshold. The merged undirected edge whose opinion value is bigger than the threshold was labeled '+', otherwise was labeled '-'.

In a given social network, in order to verify the traditional balance assumption, we should firstly determine a random baseline for the triadic closures distribution. In other word, in a social network, given a prior proportion of the positive (signed by '+') edges noted as $p_+$, and the proportion of negative (signed by '-') edges noted as $1-p_+$, the random baseline for the triadic closures distribution is the proportion of balanced triadic closures when all the positive and negative edges are randomly distributed. Assuming that the proportion of the positive edges labeled '+' is $p_+$ ($0 \leq p_+ \leq 1$). According to the traditional balance theory introduced in section 3.3.31, we can calculate the proportion of balanced triadic closures $P_{balance\_traditional}$ in random distribution using Formula (9), where $p_+^3$ is the probability of balanced tringles labeled '+++', and $3p(1-p_+)^2$ is the probability of balanced tringles labeled '+--', '-+-', and '--+'

$$P_{balance\_traditional} = p_+^3 + 3p(1-p_+)^2 \qquad (9)$$

In Figure 6, we illustrate the proportion of balanced triangles with different proportion of positive edges. In the experiments, we change the proportion of positive edges by tuning the threshold of '+/-' labeling. In Figure 6, the abscissa is the proportion of positive edges, the ordinate is the proportion of balanced triangles. The red curve is the baseline of random distribution. The other four curve are results on four language features, respectively.

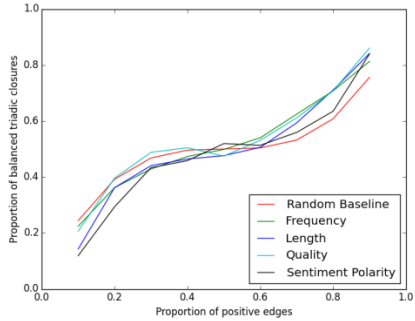

**Figure 6. Proportion of balanced triangles with different proportions of positive edges in traditional balance theory**

In Figure 6, we have two observations. First, we can intuitively find that the balanced triangles proportion and the proportion of positive edges are positively correlated. This is consistent with the fact that in traditional balance theory, two balanced tringles have more positive edges than two unbalanced triangles. Second, we can find that the proportions of balanced triangles on all four features increase faster than the random baseline with the increase of positive edges proportion. When the positive edges proportion is bigger than 0.5, the proportions of balanced triangles is greater than random baseline on all features.

From a dynamic point of view, the observations indicate that when more interrelationship turn to be positive in a social network, the social interrelationship triangles will turn to be balanced at a faster rate. From a static point of view, the balanced triangles may not always have advantages than unbalanced triangles. Its advantage is more significant in the social network where positive relationship accounted for the majority.

### 4.5.2 The verification of extended balance theory

Based on the extension of traditional balance theory in section 3.3.2, in this experiment, we investigate whether the balance assumption is still correct in the extended balance theory.

In the extended balance theory, a triadic closure is balanced when the opinions' difference on the third party and the opinions' difference on the interrelationship has the same sign. Given a threshold, we turn the above two differences into '+/-' signs. The difference greater than the threshold is signed '+', and vice versa.

We suppose the proportion of positive difference (signed by '+') is $p'_+$. In extended balance theory the tringles with '++' and '--' signs are balanced. Therefore, we calculate the proportion of balanced triadic closures $P_{balance\_extended}$ in random distribution using Formula (10).

$$P_{balance\_traditional} = p_+^2 + (1-p_+)^2 \qquad (10)$$

As in section 4.5.1, we can adjust the proportion of interrelationship with positive difference (i.e., difference bigger than threshold) with different threshold values. Then, we can investigate whether the balance assumption is still correct by investigating whether the balanced tringles are still dominant in extended balance theory.

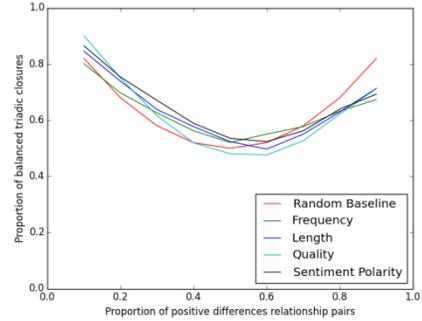

**Figure 7. Proportion of balanced triangles with different proportions of positive differences in extended balance theory**

The result curves are shown in Figure 7. The abscissa is the proportion of interrelationship pairs having positive differences, the ordinate is the proportion of balanced triadic closures. There are five curves in the diagram among which the red one represents the random baseline according to Formula (7), the red curve is the baseline of random distribution. The other four curve are results on four language features, respectively.

In Figure 7, we have two observations. First, we can find that when the proportion of the positive difference interrelationship pairs is around 0.5, the proportion of balanced triadic closures reaches the lowest point. The reason for this phenomenon is that when the proportion of positive difference pairs is equal to that of negative difference pairs, there is the maximum probability to form triadic closures with '+-' signs which is unbalanced.

Second, we also find that the proportion of the balanced triadic closure is higher than the random baseline with the increase of the proportion of negative difference relation pairs (move to the left side on X-axis). This result gives a new interpretation to the balance theory based on the asymmetric opinions measurement. The homogeneity assumption of the traditional balance theory is that in a social triadic closure, the more the two individuals' opinions on the common friends is consistent, the more positive the interrelationship between them is, and vice versa. Based on the results of this experiment, we can extend this assumption. The new assumption can be stated as: in a social triadic closure, the more the two individuals' opinions on the common friends is consistent (more negative difference on third party), the more similar the two participants' opinions on their interrelationship are, i.e., less asymmetric on bidirectional opinions (more negative difference on interrelationship).

## 4.6 Visualization of Enron Social Network

In addition, in Figure 8 and 9, we further provide a visualization of the asymmetric opinions and balanced triangles in a part of Enron email social network. Each node is an email user and the edge pair between two nodes are their directed interrelationships. The size of a node indicates the degree of the node. The thickness of the edges illustrates the language feature value on each direction of interrelationships. Therefore, the asymmetry of the opinions can be illustrated by the difference of the two edges' thickness between each nodes pair. Furthermore, the balanced triangles is colored red in both figures. Figure 8 and 9 shows the colored results by traditional balance theory and extended balance theory, respectively.

The social network visualization in Figure 8 and 9 support our above conclusions in an intuitive way. First, it can be seen that the bidirectional feature values, i.e., the bidirectional opinions on interrelationships are more symmetric and balanced in a dense community, where an individual's tend to have more friends and two individuals tend to have more common friends. This observation support the conclusions in section 4.3 and 4.4. Second, in both figures, balanced triangles (red triangles) are all in majority, which support the conclusions in section 4.5.

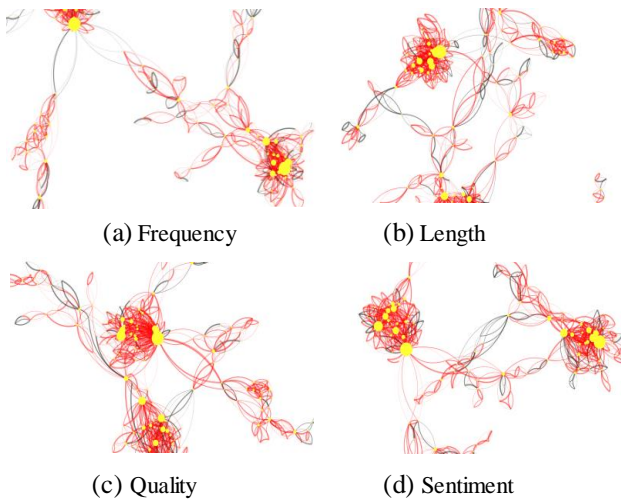

(a) Frequency  (b) Length

(c) Quality  (d) Sentiment

**Figure 8. Visualization of asymmetric opinions and traditional balanced triangles on Enron social network**

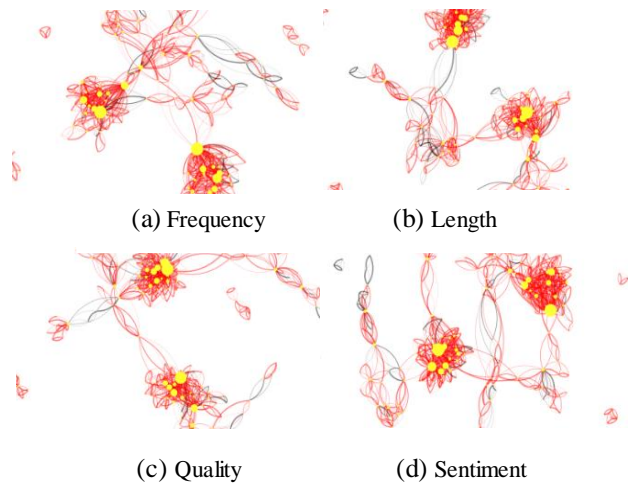

(a) Frequency  (b) Length

(c) Quality  (d) Sentiment

**Figure 9. Visualization of asymmetric opinions and extended balanced triangles on Enron social network**

## 5. CONCLUSIONS

In this work, we investigated whether the structural context features in online social network are latent factors of individuals' asymmetric opinions on their interrelationship. We characterize individuals' bidirectional opinions on interrelationships with their interactive language features following the existing method. As a case study, on Enron email dataset, we analyzed the correlations between multiple structural context features and the degree of opinions' asymmetry. The investigated structural features includes the degree and clustering coefficient of vertexes, the embeddedness of the edges and the balance of the triadic closures. During the investigation, an extension is also made to traditional balance theory to meet the form of asymmetric opinions.

With the investigation on Enron email, we find that in this case, some of the structural context features in social network can be latent factors influence the individuals' asymmetric opinions on their interrelationship. First, a more sociable person, i.e., a person with more friends, tend to be more active to cater to the partners' opinions in social interrelationships. Second, two individuals sharing more common friends tend to have more similar opinions on their interrelationship, which can be an extension of the traditional theory of social triadic closure. Third, in a social triadic closure, the two individuals' having more similar opinions on their relationship with the third person also tend to have more similar opinions on their interrelationship, which can be an extension of the traditional social balance theory. Fourth, by extending the traditional balance theory to directed social triangle, we show that directed social triangle also tend to be balanced in Enron email dataset.

These results reveal that the subjective features on a social interrelationship is not only related to individuals' own opinions but also related to their structural context in social network. This work also encourage us to further study the interaction between the topology and content in social network, and measure the social relationship synthetizing the objective and subjective features in future work.

## 6. ACKNOWLEDGMENTS

This work is supported by the Chinese National Program on Key Basic Research Project (2013CB329304), National High-tech R&D Program of China (2015AA015403) and Tianjin Younger Natural Science Foundation (14JCQNJC00400).